# Phonon Mechanism in the Most Dilute Superconductor: *n*-type SrTiO$_3$

Lev P. Gor'kov[1,2]

[1] NHMFL, Florida State University, 1800 E. Paul Dirac Drive, Tallahassee, Florida, 32310, USA
[2] L.D. Landau Institute for Theoretical Physics of the RAS, Chernogolovka, 142432, Russia

Superconductivity of doped SrTiO$_3$ that remained enigmatic for half a century is treated as a particular case of the broader concept of the non-adiabatic pairing, that is, mediated by phonons with a frequency comparable or larger the Fermi energy. Based on recent experiments we introduce the notion of the mobility edge in doped strontium titanate. Of the states above the mobility edge the electronic spectrum consists of three conduction bands filling successively with further doping. We argue that superconductivity of the doped SrTiO$_3$ results by the interaction of electrons with *several* longitudinal (LO) optical photons with a frequency *much larger* than the Fermi energy. The presence of immobile charges increases the "optical" dielectric constant compared to its value for clean SrTiO$_3$, thus reducing and controlling the interaction. Each band contributes to the Cooper instability and exhibits a superconducting gap in the energy spectrum at low temperatures. The theory predicts maxima in the dependence of the transition temperature $T_c(n_s)$ on the number of electrons $n_s$ owing to the following mechanism. Doping by the electrons increases density of states at the Fermi surface and $T_c(n_s)$ initially grows up. On the contrary, screening on the part of accumulating charges tends to reduce the electrical fields inherent in LO modes and with larger concentrations the matrix element of interaction between electrons and LO phonons decreases. The compromise between the two tendencies leads to maxima in the $T_c(n_s)$-dependence, providing interpretation to one of the most intriguing experimental findings in Xiao Lin *et al* [Phys. Rev. Lett. 112, 207002 (2014)]. Having reached a maximum in the third band, temperature of the superconducting transition finally decreases, rounding out the $T_c(n_s)$-dome, the three maxima in $T_c(n_s)$ with accompanying superconducting gaps emerging consecutively as electrons fill successive bands. This arises from attributes of the LO optical phonon pairing mechanism. The LO phonons mechanism opens the path to increasing the temperature of the superconducting transition in bulk transition-metals oxides and other polar crystals and in the two-dimensional charged layers forming at interfaces.

## I. INTRODUCTION

In the rapidly growing field of transition-metal-oxides electronics [1], SrTiO$_3$ is a key material with the unique physical properties. A broadband insulator, SrTiO$_3$ is the rare case of a paraelectric with the ferroelectric transition aborted by the quantum fluctuations and is remarkable, in particular, for a large dielectric constant that at helium temperatures reaches the enormous value of 25000 [2].

A number of features in the low temperature behavior of SrTiO$_3$ are still poorly understood, primarily its electronic properties. Perhaps most challenging is the nature of the superconductivity in doped SrTiO$_3$. Upon doping (via a chemical substitution or reducing the oxygen content), SrTiO$_3$ displays low-temperature superconductivity already at concentrations of doped electrons as low as $n_s \approx 5.5 \times 10^{17} cm^{-3}$; as such doped SrTiO$_3$ was dubbed the "most dilute superconductor" [3]. At the LaAlO$_3$/SrTiO$_3$ interfaces the electrons form a metallic layer on the SrTiO$_3$ side that displays two-dimensional (2D) ferromagnetism and superconductivity [4, 5]. Further, one unit-cell-thick layer of FeSe on the surface of doped SrTiO$_3$ becomes superconducting at record temperatures $T_C \geq 100K$ [6]. These and other examples suggest that the mechanism of superconductivity in SrTiO$_3$ might differ from that in the BCS model. Below we focus on superconductivity in *bulk* doped SrTiO$_3$ and argue that it owes its origin to interaction between electrons and longitudinal (LO) polar phonons with frequencies far exceeding the Fermi energy.

Superconductivity in SrTiO$_3$, first discovered in 1964 in reduced samples with $T_c = 0.25 \div 0.28 K$ ($n_s \approx 2.6 \times 10^{18} cm^{-3}$) [7], was supposed initially to be another case of phonon-mediated Cooper pairing generalized to the case of doped multi-valley semiconductors [8]. The limit of a small number of carriers is known to present especial challenge for theory as at low doping the Fermi energy $E_F$ is small, in fact, even smaller than the cutoff frequency $\omega_0$ that in BCS is of the order of the Debye temperature.

The dimensionless coupling parameter of the BCS theory $\lambda_{BCS}$ is proportional to the product of a matrix element of electron-phonon scattering $V_{e-ph}$ and the density of states $\nu(E_F)$ at the Fermi surface: $\lambda_{BCS} \propto V_{e-ph} \times \nu(E_F)$. While the former is of the atomic scale, as in metals, $\nu(E_F)$ is small. It was suggested in [9] that in polar semiconductors the long-range interaction with a longitudinal (LO) optical mode can compensate the smallness of density of states. After the analysis the authors, however, came to the conclusion that this mechanism may be effective *only* if the frequency $\omega_{LO}$ of such phonon is *smaller* than the Fermi energy.

Understanding of superconductivity in the doped insulator SrTiO$_3$ has met with additional difficulties (for a brief review see [10, 11]). Thus, the popular multi-valley model [8] turned out not to be relevant as in the energy spectrum of the cubic SrTiO$_3$ there is only one minimum at the $\Gamma$-point built on the titanium $3dt_{2g}$-levels [11, 12, 13]. For the same reasons not applicable is the mechanism [14] implying softening of the transverse optical phonon near the ferroelectric transition.

That the Debye temperature for strontium titanate is comparable with or, at low enough doping, even larger than $E_F$, was probably first stressed in [10]. (In [10], however, superconductivity in doped SrTiO$_3$ and the dependence of $T_c$ on concentration was described in terms of what the author called as the "plasmon-polar optical phonon mechanism". No such a mechanism exists, as will be seen from below).

Recent experiments [3, 15, and 16] demonstrated that upon doping SrTiO$_3$ passes a sequence of stages before reaching at last the metallic ground state at a high enough concentration. We show below that taking this evolution into consideration is necessary for understanding the low temperature physics in doped SrTiO$_3$. In that follows, we explore interactions between the electrons doped into the conduction band and longitudinal optical (LO) phonons. One vital difference from [9] is that there are *four* LO phonons at the $\Gamma$-point in SrTiO$_3$ (in the cubic phase [17]). The other one comes about from the fact that the presence of immobile charges embedded via the doping process significantly affects the dielectric characteristics of the material and, hence, the matrix element for the interactions of electrons with LO phonons. Most of results below are obtained analytically. Among other things, the theory predicts the appearance of maxima in the dependence $T_c(n_s)$ on the number of carriers $n_s$, providing the explanation for one of the most intriguing experimental findings [15].

Mechanisms of superconductivity in *semiconductors* were studied in [8, 9]. More often than not, superconductivity is observed in *heavily* doped semiconductors (see recent review [18]). The mechanism of optical phonons [9] combined with the traditional approach [19] to account for the Coulomb repulsion was successful in the predicting superconductivity in several multi-valley semiconductors [8].

As known, in metals the weak coupling BCS is generalized to the case of arbitrary strong electron-phonon interactions in the Migdal-Eliashberg equations [20, 21]. The value of the adiabatic parameter $\omega_0/E_F$ is not readily accessible in doped semiconductors: the Fermi energy is a function of doping and the frequencies of the relevant phonon modes is usually unknown. It is worth mentioning, in this connection, the case of diamond with $T_C$ up to $4K$ in which the Debye temperature and Fermi energy Fermi are comparable $T_D \approx E_F \sim 2000K$ [18]. (In the common BCS-like metals $T_C \sim 10^{-2}\theta_D$ where $\theta_D$ is the Debye temperature).

As the frequencies of the polar phonon modes in bulk *pure* SrTiO$_3$ are known and are high [22], electron-phonon interactions in doped SrTiO$_3$ present the *extreme* case of the *"anti-adiabatic"* limit $\omega_0 \gg E_F$. We argue that apart from the self evident significant interest in the realization of such extreme conditions specifically in the doped strontium titanate, the notion of superconductivity mediated by phonons with a frequency higher or of the same order as the Fermi energy calls for more general exploration of peculiarities inherent in the non-adiabatic pairing mechanisms.

One part of the problem concerns competition between the Coulomb repulsion and the phonon-mediated attraction. In the adiabatic regime from the condition $\omega_0/E_F \ll 1$ follows the "retardation" effect of the BCS theory: two electrons of the pair evade each other, being apart on the distances $d \sim v_F/\omega_0$ larger by the factor $E_F/\omega_0 \gg 1$ than the atomic scale $a \approx 1/p_F$. Effectively, this decreases role of the Coulomb repulsion because the latter is screened on the atomic distances.

In the opposite limit of $\omega_0/E_F \gg 1$ electrons of the pair feel both the Coulomb repulsion and the potential of the local lattice distortion instantaneously. For the Cooper pair to be formed the strength of the phonon attraction must exceed the direct Coulomb repulsion. For the general theory of electron-phonon interaction in the non-adiabatic case the immediate difficulty is that the customary mathematical apparatus of the Migdal-Eliashberg equations is not applicable at $\omega_0/E_F \geq 1$; the key advantage of the Migdal theory, the possibility of neglecting all contributions from the so called "crossing" diagrams is lost. We demonstrate below that the case of doped SrTiO$_3$ is of particular value where the most significant peculiarities proper to a non-adiabatic superconductivity can be deduced in the *logarithmic* approximation.

## II. EXCITATIONS IN INSULATING SrTiO$_3$

Consider at first two excitations in the conduction band of *pure* SrTiO$_3$ that scatter on each other via virtual exchange of LO phonons. In a polar semiconductor the scattering matrix element has the form [9]:

$$\Gamma_{GFL}(\vec{p}, \varepsilon_n | \vec{k}, \varepsilon_m) = \frac{4\pi e^2}{\kappa_\infty q^2} - \frac{4\pi e^2}{q^2}\left(\frac{1}{\kappa_\infty} - \frac{1}{\kappa_0}\right) \times \frac{\omega_{LO}^2}{\omega_{LO}^2 + (\varepsilon_n - \varepsilon_m)^2}. \quad (1)$$

We apply the thermodynamic technique (see [23]) that is more convenient for the analysis of the Cooper instability compared to the traditional method in the literature of solving the linear integral equation for the gap parameter on the real frequency axis, as this technique circumvents unnecessary mathematical difficulties arising from the pole singularities in the kernel of the integral equation.

In (1) $\omega_{LO}^2$ is the square of the LO phonon frequency; $\vec{q} = \vec{p} - \vec{k}$ and $\varepsilon_n - \varepsilon_m$ are the momentum and frequency the electrons exchange upon scattering; $\kappa_0$ and $\kappa_\infty$ are the static and optical dielectric constants, respectively. If the frequency of LO phonon is large $\omega_{LO} \gg E_F$, at low temperatures one can omit $\varepsilon_n - \varepsilon_m$ in the denominator of the phonon Green function. Doing so in Eq. (1), one obtains:

$$\Gamma_{GFL}(\vec{p}, \varepsilon_n | \vec{k}, \varepsilon_m) = \Gamma_{GFL}(q) = \frac{4\pi e^2}{\kappa_0 q^2} > 0. \quad (2)$$

That is, the initial Hamiltonian (2) [9] corresponds to the *repulsion* between electrons and superconductivity may arise only, as was pointed out by the authors, if repulsion is overcome via the retarded action of the phonons as in the BCS. For that, however, the phonon frequency $\omega_{LO}$ must be small compared to the Fermi energy.

As mentioned, in case of SrTiO$_3$ there are *four* LO phonons at the $\Gamma$-point [17]. Strictly speaking, matrix elements of the interaction between electrons and LO phonons in a *multi-mode* polar crystal have a more complicated form than in Eq. (1). Characteristic for SrTiO$_3$, however, one phonon mode has a giant LO-TO gap [17]. For such LO mode Eq. (1) is correct, at least approximately. As in [9], its contribution practically compensates the direct Coulomb term, as the static dielectric constant is large ($\kappa_0 \approx 2 \times 10^4$ [2]). Of the three LO phonons one mode is not infra-red active [17]. As the result, instead of (2), one has:

$$\Gamma(q) = \frac{4\pi e^2}{\kappa_0 q^2} - \alpha_{eff}^2 \frac{4\pi e^2}{q^2}\left(\frac{1}{\kappa_\infty} - \frac{1}{\kappa_0}\right) \approx -\alpha_{eff}^2 \frac{4\pi e^2}{q^2 \kappa_\infty} < 0. \quad (3)$$

Here the factor $\alpha_{eff}^2$ comes about from the interaction of electrons with the two remaining LO-polar phonons. In the *clean* strontium titanate their contribution into (3) would guarantee the attractive sign of the interaction between two excitations. (Formation of polarons and other polaron-related effects are not discussed. The expression in (3) was further simplified by taking into account that at low temperatures in SrTiO$_3$ $\kappa_\infty \ll \kappa_0$ [2]). Repeat again that Eq. (3) is the matrix element for scattering of two electrons in the conduction band of *insulating* SrTiO$_3$.

### III. DOPING SrTiO$_3$: EXPERIMENTAL SUMMARY

Recent experiments [3, 15 and 16] revealed a number of new features as concerns the microscopic properties of *doped* SrTiO$_3$. It is worth to summarize briefly some facts most essential for the discussion below.

1) Pure SrTiO$_3$ is on the verge of the ferroelectric transition with the static dielectric constant $\kappa_0$ at low-temperature $\kappa_0 \simeq 25000$ being four orders of magnitude larger that of the vacuum. The Bohr radius is very large $a_B^* = 0.53\kappa_0(m_e/m)\times 10^{-8} cm$ giving $a_B^* \approx 5 \div 7 \times 10^{-5} cm$ [2].

2) Correspondingly, the insulator-to-metal transition into the emerging "impurity band" occurs, according to the famous Mott criterion $n_s^{1/3} a_B^* > 0.26$, already at low doping $n_s^{MT} < 10^{12} cm^{-3}$. Experimentally, the crossover from the regime of the "impurity band" transport to the regime of conductivity of the coherent carriers in the conduction band starts at $n_s^* \approx 2\times 10^{16} cm^{-3}$ [16]. In order of magnitude $n_s^*$ may be taken as an estimate for the mobility edge.

Shubnikov-de Haas (SdH) quantum oscillations (QO) in the magneto-resistance are first seen at $n_s \approx 4\times 10^{17} cm^{-3}$; QO *and* superconductivity are observable in samples with somewhat higher carrier concentration $n_s \approx 5.5\times 10^{17} cm^{-3}$. At such concentrations the Fermi energy is extremely small ($E_F \approx 1.1 \div 1.3 meV$) [3, 15].

3) With $n_s$ varying between $n_s \approx 5.5\times 10^{17} cm^{-3}$ and $n_s \approx 4\times 10^{20} cm^{-3}$ temperature of the superconducting transition $T_c$ in [15] varies from $\approx 0.07 K$ to $\approx 0.4 K$.

4) In the experimental dependence of $T_c$ on the dopants concentration [15] are distinguishable several regimes. First, $T_c$ increases ($n_s$ varies from $n_s \approx 5.5\times 10^{17} cm^{-3}$ to $n_s \approx 1.05\times 10^{18} cm^{-3}$). After reaching a maximum at $n_{s\max} \approx 2\times 10^{18} cm^{-3}$ $T_c$ then *decreases* before to start growing again as electrons begin to fill the next band. Observation of a new SdH frequency at $n_{c1} \simeq 1.2\times 10^{18} cm^{-3}$ signifies the concentration at which the chemical potential touches the bottom of the second band. At last, another frequency of QO above $n_{s2} \approx 2\times 10^{19} cm^{-3}$ signals that the chemical potential has reached the bottom of the third band [15].

From this behavior one can infer, in particular, that the threefold degeneracy of the spectrum of the cubic phase of SrTiO$_3$ at the $\Gamma$-point is lifted as the growth of the quantum fluctuations is pinned in the presence of the doping centers, allowing local (tetragonal) lattice distortions.

Maxima in $T_c(n_s)$ is the most remarkable feature among the findings [3, 15]. It is argued that such maxima are the signature of LO polar phonons that constitute the mechanism of the superconductivity of doped SrTiO$_3$.

## IV. INTERACTION OF ELECTRONS WITH LO PHONONS IN DOPED SrTiO$_3$

The Hamiltonian (3) describes the interaction between electrons via the exchange by LO phonons in *clean* SrTiO$_3$. At least at low doping, like in a semiconductor the coupling constant for exchange by ordinary (acoustic) phonons is small, and one has to revisit the conjecture that the electron-lattice interaction in *doped* SrTiO$_3$ is long-ranged and is governed by the electric fields inherent in LO optical modes. In the Fröhlich Hamiltonian:

$$\vec{P} = F_C \vec{u}, \quad (4)$$

for the connection between the polarization $\vec{P}$ and the lattice displacement $\vec{u}$ one must however take into account possibility of screening mechanisms.

Details of these mechanisms are not *a priori* clear in the case in hand. In regard to the Mott criterion $n_s^{1/3} a_B^* > 0.26$ for the transition into a "metallic" impurity band, we note, together with [16], that the binding energy $E_B = -13.6 eV \times (m/m_e) \kappa_0^{-2}$ of a hydrogen-like center at $\kappa_0 \simeq 24000$ is so small that doping centers must be ionized at any temperature of interest. One obvious consequence of that is the absence of a distinct gap separating the "impurity band" from "the bottom" of the conduction band. Instead, there can be a threshold in the density of states equivalent, in effect, to *the mobility edge*. As mentioned above, the crossover towards the coherent conduction band transport is observable above concentration $n_s^* \approx 2 \times 10^{16} cm^{-3}$ [16].

Lattice displacements in a polar crystal are accompanied by formation of electric dipoles also in the presence of a disorder. Therefore, as before, several optical phonons propagating in doped SrTiO$_3$ carry over longitudinal electric fields. However, the *metallic* regime experimentally firmly establishes itself at a concentration more than a factor ten higher $n_s^* \approx 2 \times 10^{16} cm^{-3}$. We argue that the concept of the Fröhlich Hamiltonian for electrons in the *conduction band* of *doped* SrTiO$_3$ at such concentrations must be explored anew.

We **hypothesize** that the carriers with their energy *below* the mobility edge contribute mainly by changing the value of the "optical" dielectric constant. The new Fröhlich matrix element (4) for electrons in the conduction band will be written down in the same canonical form Eq. (1):

$$F_C = \left[4\pi e^2 \frac{\hbar\omega_{LO}}{2}\left(\frac{1}{\bar{\kappa}_\infty} - \frac{1}{\bar{\kappa}_0}\right)\right]^{1/2}, \qquad (5)$$

where $\bar{\kappa}_\infty$ and $\bar{\kappa}_0$ in (5) ($\bar{\kappa}_0 \gg \bar{\kappa}_\infty$) are the optical and the static dielectric constants *in the presence of localized electrons* in the background.

One hydrogen-like center possesses the dipole moment $\approx (9/2)a_B^{*3}$. As $a_B^*$ is large, reaching $a_B^* = 5 \div 7 \times 10^{-5} cm$, local centers can significantly contribute to the polarizability. At $n_s \approx n_s^* \approx 2\times 10^{16} cm^{-3}$ that may results in a large dielectric constant $\bar{\kappa}_\infty \sim 10^2 \div 10^3$.

The Coulomb potentials can also be screened by the mobile carriers in the conduction band. Combining the two mechanisms, instead of Eq. (3) one writes:

$$\tilde{\Gamma}(q,\omega_{mn}) = \frac{4\pi e^2}{Q^2(q,\omega_{mn})} \times \left[\frac{1}{\bar{\kappa}_0} - \alpha_{eff}^2\left(\frac{1}{\bar{\kappa}_\infty} - \frac{1}{\bar{\kappa}_0}\right)\right] \approx -\alpha_{eff}^2 \frac{4\pi e^2}{Q^2(q,\omega_{mn})\bar{\kappa}_\infty} < 0. \qquad (6)$$

Eq. (6) finalizes our choice of the model. Both $\bar{\kappa}_\infty$ and $\alpha_{eff}^2$ are model parameters that may depend on sample quality. It is convenient to redefine $\bar{\kappa}_\infty$ in (6) by taking $\alpha_{eff}^2 = 2$. In the single band with the parabolic energy spectrum $\varepsilon(\vec{p}) = p^2/2m$ doped electrons fill all states up to the chemical potential $\mu \equiv E_F = p_F^2/2m$. (The case of several bands is considered later).

In the so called Random Phase Approximation (RPA) the general form of $Q^2(q,\omega_{mn})$ in (6) is:

$$Q^2(q,\omega_{mn}) = q^2 + \kappa_{TF}^2 S(q,\omega_{mn}). \qquad (7)$$

Here

$$S(q,\omega_{mn}) = \int_0^1 \frac{(\vec{v}_F \cdot \vec{q})^2 d\mu}{(\vec{v}_F \cdot \vec{q})^2 + \omega_{mn}^2}, \qquad (7a)$$

and $\kappa_{TF}^2$ is the square of the Thomas-Fermi radius:

$$\kappa_{TF}^2 = (4e^2 m p_F / \bar{\kappa}_\infty \pi \hbar^3). \qquad (7b)$$

It will be shown later that at the calculations below one can assume in (6, 7a) $\omega_{nm} = 0$. That is, $S(q,0) = 1$ (see Appendix I).

Introduce the notation $\bar{a}_B = \bar{\kappa}_\infty \hbar^2 / e^2 m$ for the "optical" Bohr radius and rewrite $\kappa_{TF}^2 = (4e^2 m p_F / \bar{\kappa}_\infty \pi \hbar^3) = 4p_F^2 / (\pi \bar{a}_B p_F / \hbar)$. In passing, note that from the formal point of view the regime of the large Mott parameter $n_s^{1/3} a_B^* \gg 1$ is analogous to the regime of the "degenerate plasma". Owing to the mobility edge and the presence of immobile carriers applicability of RPA in the conduction band is controlled by the value of $\pi p_F \bar{a}_B / \hbar \gg 1$.

## V. COOPER INSTABILITY IN LOGARITHMIC APPROXIMATION

For ordinary superconductors the Migdal adiabatic provision $\omega_0 / E_F \ll 1$ allows simplifying the diagrammatic expansion for the scattering amplitude by omitting all the so called "crossing diagrams". The result is reduced to the closed system of the Migdal-Eliashberg equations [20, 21]. In non-adiabatic case this theoretical tool is lost. Fortunately, the basic features of the superconductivity of *doped* SrTiO$_3$ can be explored taking advantage of the logarithmic approach. The latter is applicable at the small $T_C \ll E_F$ and, as mentioned above, in the case in hand $T_C = (10^{-3} \div 10^{-2}) E_F$.

It is helpful to briefly enumerate main steps at the derivation of the linear integral equation for the superconducting order parameter. Superconductivity manifests itself in the occurrence at $T = T_C$ of the pole in the scattering amplitude $\Gamma(p, q-p | p', q-p')$ at zero total momentum and frequency $q = 0$ [23]. The exact amplitude is the sum of all diagrams in the Cooper channel. Denoting $\Gamma(p, q-p | p', q-p')|_{q=o} \equiv \Gamma(p | p')$, the equation for $\Gamma(p | p')$ reads:

$$\Gamma(p, | p') = \tilde{\Gamma}(p, | p') - \frac{T}{(2\pi)^3} \Sigma_{n'} \int d\vec{k} \tilde{\Gamma}(p, | k) G(k) G(-k) \Gamma(k, | p') . \quad (8)$$

In practice, instead calculating $\Gamma(p | p')$ from Eq. (8), the transition temperature is determined via the eigenvalue of the homogeneous equation for the gap function $\Psi(p)$:

$$\Psi(p) = -T \Sigma_m \int \frac{d\vec{k}}{(2\pi)^d} \tilde{\Gamma}(p | k) \Pi(k) \Psi(k) . \quad (9)$$

The Cooper instability owes its origin to the logarithmic divergence in the blocks $\Pi(k)$ represented in Eq. (8) by the product of the two Green functions $G(k)G(-k)$. In the thermodynamic formulation $G(k) = [iv_n - (\vec{k}^2 - p_F^2)/2m]^{-1}$ and one has:

$$\Pi(k) = G(k)G(-k) = \frac{1}{v_m^2 + [(\vec{k}^2 - p_F^2)/2m]^2} . \quad (10)$$

Solving the integral equation (9) would give for $T_c$ a BCS-like weak-coupling form:

$$T_c = W \exp(-1/\lambda). \quad (11)$$

Instead, we determine the dimensionless constant $\lambda$ in (11) *exacltly without solving* Eq. (9). As to the prefactor $W$, its value will be known only to the accuracy of a numerical factor of the order of the unity.

Substituting (10) in the right hand side of equation (9) gives:

$$-\frac{T}{(2\pi)^3} \sum_{n'} \int d\vec{k}\, \tilde{\Gamma}(p|k) G(k) G(-k) \Psi(k)$$

$$\Rightarrow -T \sum_{n'} \int \tilde{\Gamma}(p|k)[mp_F \sin\theta d\theta)/(2\pi)^2] d\varsigma \frac{1}{v_m^2 + \varsigma^2} \Psi(k). \quad (12)$$

(In (12) $\varsigma = (\vec{k}^2 - p_F^2)/2m \approx v_F(p - p_F)$; $\theta$ is the angle between two vectors $\vec{p}$ and $\vec{k}$).

Let the vectors $\vec{p}$ and $\vec{k}$ in (12) be *on the Fermi surface*. The expression for the vertex $\tilde{\Gamma}(p|k) \equiv \tilde{\Gamma}(\theta)|_{FS}$ stands in front of the logarithmic singularity in Eq. (12):

$$\left[\int_0^\pi \sin\theta d\theta \tilde{\Gamma}(\theta)|_{FS}\right] \frac{mp_F}{(2\pi)^2} \times \int_0^W \frac{d\varsigma}{\varsigma} \text{th}\frac{\varsigma}{2T} \Rightarrow \lambda \ln\left(\frac{2W\gamma}{\pi T}\right). \quad (13)$$

Thereby, Eq. (13) gives for $\lambda$ the definition:

$$\lambda = \left[\int_0^\pi \sin\theta d\theta \tilde{\Gamma}(\theta)|_{FS}\right] \frac{mp_F}{(2\pi)^2}. \quad (14)$$

In (13) $W$ is a characteristic scale in the dependence of $\tilde{\Gamma}(p|k)$ on the energy variable $\varsigma$ that plays the role of an order-of-magnitude cutoff parameter in the integral over $\varsigma$ in Eq. (12). (To determine the *exact* value of the numerical factor in the expression $T_c = const \times E_F (2\gamma/\pi) \exp(-1/\lambda)$ one must be able to solve the integral equation (9) explicitly).

## VI. EXPRESSIONS FOR THE TRANSITION TEMPERATURE

In Eqs. (13, 14) take for $\tilde{\Gamma}(p|k)$ its expression $\tilde{\Gamma}(q, \omega_{mn})$ from Eqs. (6, 7):

$$\int \tilde{\Gamma}(p|k)[mp_F \sin\theta d\theta)/(2\pi)^2 = \left[\int_0^\pi \frac{e^2 m \sin\theta d\theta}{\pi p_F[1 - \cos\theta + 2(\varsigma_p + \varsigma_k)/E_F + \kappa_{TF}^2/2p_F^2]\bar{\kappa}_\infty}\right], \quad (15)$$

substitute $\kappa_{TF}^2 = (4e^2 mp_F/\bar{\kappa}_\infty \pi \hbar^3) = 4p_F^2/(\pi \bar{a}_B p_F/\hbar)$ and integrate over $\cos\theta$. In the integral equation (9) one finally arrives to the following expression of the kernel:

$$\frac{\hbar}{\pi p_F \bar{a}_B} \ln \frac{1+(\varsigma_p +\varsigma_k)/E_F +\hbar/\pi p_F \bar{a}_B}{(\varsigma_p +\varsigma_k)/E_F +\hbar/\pi p_F \bar{a}_B}. \quad (15a)$$

(In (15) $\varsigma_k = v_F(k-p_F)$ and $\varsigma_p = v_F(p-p_F)$).

At the Fermi surface $\varsigma_k$ and $\varsigma_p$ equal zero; from (15a) follows:

$$\lambda = \frac{\hbar}{\pi p_F \bar{a}_B}\ln(1+\pi p_F \bar{a}_B/\hbar). \quad (16)$$

Expressions (15, 15a) guarantee the convergence of the integral over $\varsigma_k = v_F(k-p_F)$ in Eq. (13) at $\varsigma_k$ of the order of $E_F$. (The *exact* solution of the integral equation (9) the function $\Psi(p)$ will depend on $\varsigma_p$ as well). In other words, since the phonon frequency is much larger the Fermi energy, the latter becomes the only pertinent energy scale in the problem. The cutoff parameter in Eq. (13) is $W \approx E_F$. Expression for $T_C$ Eq. (11) can be presented in the following convenient form:

$$T_C = const \times \frac{\gamma}{\pi^3} \times \frac{\hbar^2}{m\bar{a}_B^2}\left[\bar{T}_1(\pi p_F \bar{a}_B/\hbar)\right]. \quad (17)$$

Denoting in (17) $x = \pi p_F \bar{a}_B/\hbar$; the function $\bar{T}_1(x) = x^2 \exp[-x/\ln(1+x)]$ is plotted in Fig.1.

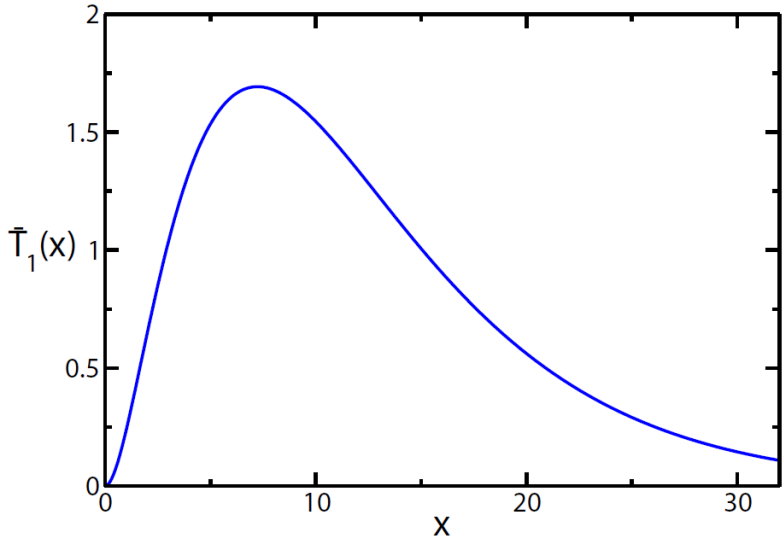

FIG.1. (Color online) Temperature of the superconducting transition $T_C$ as function of the concentration of electrons $n_s$ (in the first band). $T_C(n_s)$ is proportional to $\bar{T}_1(x) = x^2 \exp[-x/\ln(1+x)]$. The relation between the variable $x$ and the concentration is $x = \pi p_F \bar{a}_B/\hbar \simeq 5.74 \times (n_s \times 10^{-18})^{1/3}$ (see text). The function $\bar{T}_1(x)$ has maximum at $x \simeq 7.18$. At larger $x$ $\bar{T}_1(x)$ tends to zero and, as discussed in the text below, at higher concentrations superconductivity stems from filling the second band and then, at even

higher $n_s$, from that in the third one. The emerging maximum in the $T_c(n_s)$-dependence has been observed experimentally [15].

To test the relevance of the theoretical expression (17) to the experiment, consider $T_c(n_s)$ in the vicinity of the maximum in Fig.1b [15]; the latter is reached at $n_{s\max} \approx 2 \times 10^{18} cm^{-3}$ ($T_{C\max} \approx 0.2K$). (Contributions into $T_c(n_s)$ from the second band at these concentrations supposedly remain small). The maximum of $\bar{T}_1(x)$ in Fig.1 $\bar{T}_1(x_{\max}) \simeq 1.69$ is at $x_{\max} \simeq 7.18$.

Substituting $p_F / \hbar \equiv (n_s)^{1/3}(3\pi^2)^{1/3}$ into $(\pi p_F \bar{a}_B / \hbar)_{\max} \simeq 7.18$, one finds $\bar{a}_B \approx 0.58 \times 10^{-6} cm$. According to Fig. 4a [15], the mass of the lower band is $m_1 \approx 1.8 m_e$. Eq. (17) gives for $T_{C\max} \approx const \times 1.1K$. Thereby, $const \approx 1/5$. ($\bar{a}_B = 0.53 \bar{\kappa}_\infty (m_e / m_1) \times 10^{-8} cm \approx 0.58 \times 10^{-6} cm$ allows the estimate for the dielectric constant $\bar{\kappa}_\infty \approx 2 \times 10^2$; for the pure bulk SrTiO$_3$ $\kappa_\infty = 5.2$ [2]). (Note the large $\pi p_F \bar{a}_B / \hbar \geq 7$, as it was expected,).

Eq. (17) suggests the simple physical explanation for the maximum of $T_c(n_s)$ in Fig.1b [15]. At the start of the doping $T_c$ in Fig.1 increases, but at higher concentrations important becomes the RPA screening and $T_c$ decreases as the Thomas-Fermi radius Eq. (7) $r_{TF} \sim 1/\kappa_{TF} \propto n_s^{-1/3}$ gets shorter.

So far only one band with the isotropic energy spectrum was considered in the above. As it was already mentioned earlier, at lower temperatures the lattice deformations split the threefold degenerate minimum at the $\Gamma$-point of the cubic SrTiO$_3$ into three bands [11, 12, 15 and 24]. For a single band from Eq. (17) and Fig.1 it follows that at large $n_s$ $T_c(n_s)$ would tend to zero. Therefore, at the increase of the dopant concentration superconductivity comes about with filling up the second band and then, at even higher concentrations, the third one.

The maximum is inherent in the kernel (15) and in the expression for $T_c(n_s)$ Eq. (17) and is the hall-mark of the interaction of electrons with LO phonons.

It was pointed out in the literature that some bands may be anisotropic [11, 13 and 25]. By itself, a change in the $T_c(n_s)$-behavior, as the chemical potential goes over *from one band into the other*, is a qualitative feature that seems rather insensitive to the anisotropy details. The randomness introduced in the doping process smears effects of anisotropy. In fact, the data on QO confirm that the Fermi surfaces are less anisotropic than expected [13, 15, and 26]. For these reasons the doping dependence of $T_c(n_s)$ is studied below analytically in the model of three parabolic bands with three different masses, as shown in Fig.2 a, b.

(a)

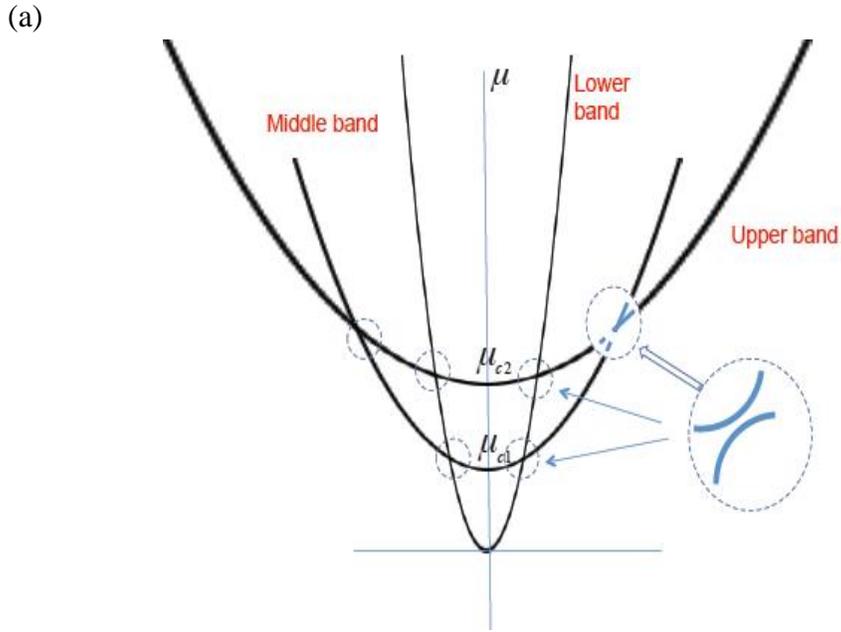

(b)

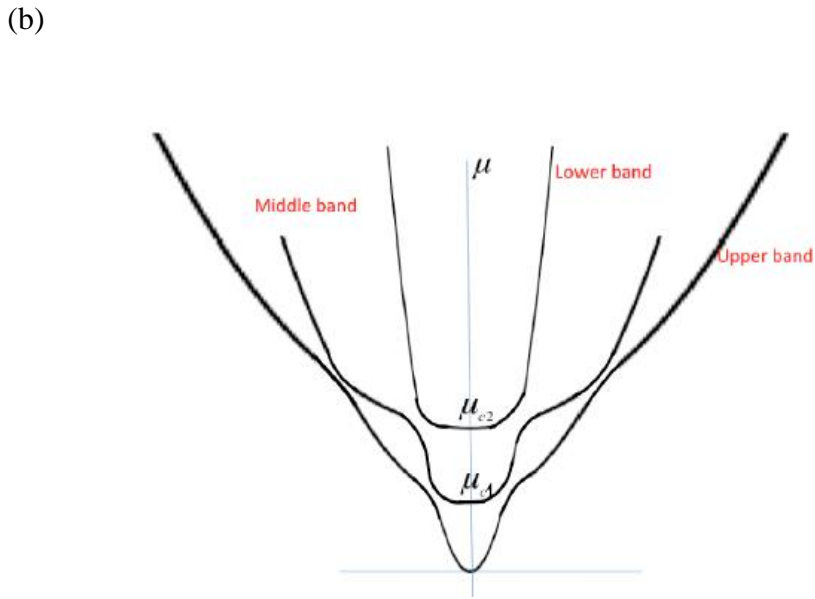

FIG. 2. (Color online) Three-band model. Top: Shown schematically, are three bands arising at low temperatures from splitting of the threefold degenerate minimum at the $\Gamma$-point in cubic SrTiO$_3$. Here $\mu_{c1}$ and $\mu_{c2}$ mark the bottoms of the Middle and of the Upper bands at the concentrations $n_{c1} = 1.2 \times 10^{18} cm^{-3}$ and $n_{c2} \simeq 2.5 \times 10^{19} cm^{-3}$, respectively [15]. As depicted, the six points of the bands intersections are marked by small dashed ovals. The degeneracy at each such intersection is lifted; the two bands locally repulse each other (see in the larger oval on the right). Bottom: The resulting pattern of the three-band spectrum (schematic). With the chemical potential moving upwards, one, two and three different frequencies in the spectrum of quantum oscillations emerge in the consecutive order. The spectrum of oscillations becomes complex when the chemical potential position falls too close to one of the intersections not far from either $\mu_{c1}$ or $\mu_{c2}$.

The new frequency emerging in the Shubnikov-de Haas oscillations at the concentration $n_{c1} \simeq 1.2 \times 10^{18} cm^{-3}$ [15] signifies reaching the bottom of the second (the Middle) bond. The derivation of Eq. (16) can be repeated, this time for the chemical potential $\mu$ in a position above the bottom $\mu_{c1} \equiv p_{c1}^2/2m_1$ of the second band. With the notations $m_2$ and $p_{F2} = (m_2/m_1)^{1/2}(p_{F1}^2 - p_{c1}^2)^{1/2}$ for the mass and the Fermi momentum in the second band, one writes $Q^2(q,0) = q^2 + (4e^2/\pi\kappa_\infty\hbar^3)[m_1 p_{F1} + m_2 p_{F2}]$. Simple calculations give the factor $\lambda_2$ in the exponent of Eq. (13) for the temperature of the Cooper instability in the second (the Middle) band ($\lambda_1 \to \lambda_2$):

$$\lambda_2 = e^2(m_2/\pi p_{F2}\bar{\kappa}_\infty)\ln\{1 + (p_{F2}^2\pi\kappa_\infty\hbar^3)/e^2[(m_1 p_{F1} + m_2 p_{F2})]\} \quad (18)$$

The relation between concentration and position of the chemical potential $\mu = p_{F1}^2/2m_1$ in the second band is:

$$[p_{F1}^3 + p_{F2}^3](1/3\pi^2\hbar^3) = n_s. \quad (19)$$

At concentrations $n_s$ above $n_{c2} \simeq 2.5 \times 10^{19} cm^{-3}$ electrons begin filling the third (the Upper) band [15]. Accordingly,

$$\lambda_3 = e^2(m_3/\pi p_{F3}\bar{\kappa}_\infty)\ln\{1 + (p_{F3}^2\pi\kappa_\infty\hbar^3)/e^2[(m_1 p_{F1} + m_2 p_{F2} + m_3 p_{F3})]\}. \quad (20)$$

Similarly to Eq. (19)

$$[p_{F1}^3 + p_{F2}^3 + p_{F3}^3](1/3\pi^2\hbar^3) = n_s. \quad (21)$$

Eqs. (16, 18-21) present the exponents $\lambda_{1,2,3}$ in expressions for the onset temperatures of superconductivity in each of the three bands when the chemical potential moves from one band to the other with increasing dopants concentration.

## VII. DISCUSSION AND CONCLUSIONS

The two expressions (18, 20) are similar to Eq. (16) in that $T_{c2}(n_s)$ and $T_{c3}(n_s)$ both in the Middle and the Upper bands display maxima at some $n_{1\max}$ and $n_{2\max}$. This stems from that $\lambda_2$ and $\lambda_3$ is small at small $p_{F2}$ and $p_{F3}$; and decrease again as the screening radius $r_{TF} \sim 1/\kappa_{TF} \propto n_s^{-1/3}$ is getting shorter at larger concentrations. In Fig.3a $\lambda_1(x)$ (Eq. (16)), $\lambda_2(x)$ (Eq. (18)), and $\lambda_3(x)$ (Eq. (20)) in each band are plotted as functions of $x = (\pi p_F \bar{a}_B / \hbar)$; the larger is the value of $\lambda$, the higher is $T_c$. (For the explicit form of relations between $x$ and the dopants concentration $n_s$, see Appendix II)

Temperatures of the transition $T_{C;1,2,3}$ in all three bands can be presented in the form

$$T_{C;1,2,3}(n_s) = (const)_{1,2,3} \times \frac{\gamma}{\pi^3} \times \frac{\hbar^2}{m\bar{a}_B^2} \times \bar{T}_{1,2,3}(x). \quad (22)$$

The dimensionless $\bar{T}_{1,2,3}(x)$ are plotted in Fig.3b. (For the analytical form of dependence of each of $\lambda_1(x)$, $\lambda_3(x)$ and $\bar{T}_{1,2,3}(x)$ on $x$, see Appendix II).

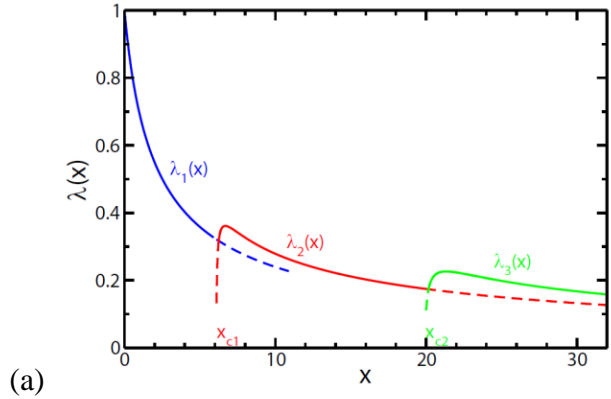

(a)

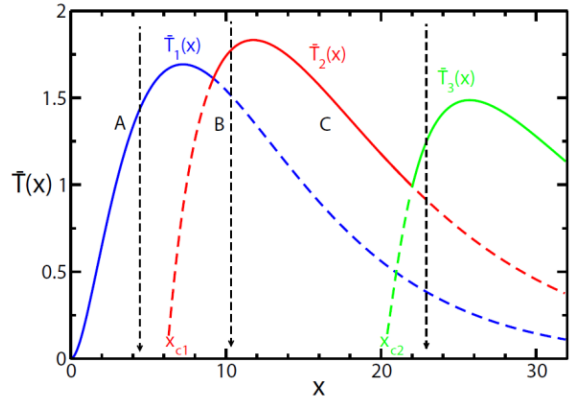

(b)

FIG.3. (Color online) Low temperature phase diagram of doped SrTiO$_3$ in the $(T, n_s)$-plane in the three band model Fig.2. (a) The concentration dependence of the $\lambda$-factors in exponents of the expression for the transition temperature $T_c = const \times W \exp(-1/\lambda)$ for each band. Relations between $x$ and $n_s$ in different bands are: 1) the Lower band at $0 < x < x_{c1}: x = \pi p_F \bar{a}_B / \hbar \simeq 5.74(n_s \times 10^{-18})^{1/3}$; 2) in the Middle band at $x_{c1} < x < x_{c2}: x^3 + (m_2/m_1)^{3/2}(x^2 - x_{c1}^2)^{3/2} \simeq 195(n_s \times 10^{-18})$; 3) in the Upper band at $x > x_{c2}$): $x^3 + (m_2/m_1)^{3/2}(x^2 - x_{c1}^2)^{3/2} + (m_3/m_1)^{3/2}(x^2 - x_{c2}^2)^{3/2} \simeq 195(n_s \times 10^{-18})$. For masses are taken their experimental values $m_1 \simeq 1.8 m_e$, $m_2 \simeq 3.5 m_e$ [15]; $m_3 \simeq 6 m_e$ [12, 15, 25].

Here and below $x_{c1} \approx 6.1$ and $x_{c2} \approx 20$ correspond to the two concentrations $n_{c1} = 1.2 \times 10^{18} cm^{-3}$ and $n_{c2} \simeq 2.5 \times 10^{19} cm^{-3}$ [15].

(b) Dimensionless temperature $\bar{T}(x)$ (for the definition, see Eq. (22) of the main text) as a function of the concentration. Full lines depict dependence on the concentration of the temperature of superconducting transition across the whole $(T, x)$-plane. Dashed lines are lines of the pairing instability in preceding bands. In principle, at a given $x$ superconductivity sets in simultaneously for all bands owing the proximity effects (non-zero inter-band matrix elements), but, as the effect of proximity is weak in the case of the pairing mechanisms under the discussion, superconductivity in these bands manifests itself mainly below these dashed lines thereby leading at low temperatures to the pattern of the well-distinguishable superconducting gaps. The three vertical arrows in (b) show that the gaps' number may vary from one single gap (A) to two gaps (B) and three gaps (C).

With no pretense to describing quantitatively the experiment[15] the three plots of $\bar{T}_{1,2,3}(x)$ in Fig.3 correctly reproduce the expected overall behavior of $T_c$ in the $(T_c, n_s)$-phase diagram of doped SrTiO$_3$, including, in particular, the appearance of the three $T_c$-maxima. Two maxima are seen in Fig. 1b [15] at $\approx 0.2 K$ and at $\approx 0.4 K$ (data at the intermediate concentrations in [15] are absent).

Two interesting qualitative predictions follow. First, with the electrons filling successively one band after another, the tunneling experiments at low temperatures are expected to reveal one, two and even three superconducting gaps developing in parallel with the increase in concentration, as shown schematically in Fig. 3b. (The two-gap structure in the $I(V)$-characteristics has been observed in [27].

Secondly, the three bands fully exhaust the electronic spectrum of SrTiO$_3$ at the $\Gamma$-point. Therefore, the $T_c$ maximum observed after electrons began filling up the third band imposes the upper limit on the value of the temperature of the superconducting transition.

In the model of three *parabolic* bands with the three different masses shown in Fig. 2 the frequencies of QO in the each band would smoothly grow proportional to the chemical potential, except a vicinity of the bands intersection. (In [15] the irregular features in QO *near* the intersection of the first (lower) band with the second (the Middle) band were interpreted as arising due to changes in the spectrum of the *first* band (see Fig.4d, e [15])).

Note in passing that the $T_c(n_s)$-dependence, asymmetric with respect to the maximum at $0.2K$ in Fig.1b [15], is reproduced poorly in the single-band picture Eq. (18) because of the closeness to the two band intersection. In fact, in Fig.1 maximum $\bar{T}_1(x)$ is at $x_{max} \approx 7.2$, but one can see in Fig. 3a that the contribution of the second band at such concentrations is already important. We do not stay on further details because of the idealized character of the model.

Finally, it is worth of emphasizing again that the notion of superconductivity mediated by phonons with the frequency exceeding the Fermi energy is different compared with the BCS where the effect of retardation in the phonon-mediated attraction at $\omega_0 \ll E_F$ circumvents the difficulty with the direct Coulomb repulsion. Notable, superconductivity in doped SrTiO$_3$ is the particular case when calculations could be done analytically.

Pairing in a non-adiabatic regime opens a new prospect, as concerns the possibility of increasing the temperature of the superconducting transition in transition-metals oxides and other polar crystals. Indeed, in the frameworks of the weak-coupling expression for $T_C$, even assuming the same value of the dimensionless coupling constant, the temperature of superconducting transition $T_c = const \times W \exp(-1/\lambda)$ now depends on the different prefactor $W \propto E_F$. This possibility, however, remains basically unexplored, as solving the integral equation (9) for the gap parameter beyond the logarithmic approximation is complicated by several additional factors. (Specifically, the expression Eq. (10) of the block $\Pi(k)$ in (9) must be modified $\Pi(k) \Rightarrow G(k)G(-k) - G_0(k)G_0(-k)$, where $G_0(k)$ is the Green function of non-interacting fermions; see in [28]).

*In conclusion,* the notion of the mobility edge is introduced as the fundamental concept underlying low temperature properties of the $n$ – doped strontium titanate. Attention is drawn to the fact that electrons doped in SrTiO$_3$ fall into the two groups with different properties. The first group consists of the localized electrons. Owing to large dielectric constant $\kappa_0 \approx 10^4$ of insulating SrTiO$_3$ these electrons occupying states below the mobility edge are responsible for large $\bar{\kappa}_\infty$ of the order from $10^2$ to $10^3$. Note that $\bar{\kappa}_\infty$ is the only free parameter in the theory.

At concentrations exceeding the threshold corresponding to the mobility edge, the superconducting pairing in the metallic bands is mediated by the interactions with LO polar phonons. Arguments given that in SrTiO$_3$ such interaction is attractive. The value of the Migdal parameter $\omega_0/E_F$ is inverted; the Fermi energy is much smaller than frequencies of the optical phonons.

With accuracy of a numeric factor temperature of the superconducting transition $T_c(n_s)$ and its dependence on the concentration $n_s$ was obtained analytically. Three maxima in $T_c(n_s)$ and three superconducting gaps emerging in succession, while electrons fill one band after another, are the hall-marks of the LO optical phonons pairing mechanism.

Estimates obtained with the help of Eq. (22) and plots in Fig. 3 suggest that the superconducting transition temperature in the electron-doped $SrTiO_3$ may vary with concentration in the interval between few hundredths to few Kelvin degrees.

The value of $\bar{\kappa}_\infty$ depends on sample quality and so does the overall pattern for superconductivity in the (T, x)-plane, as is confirmed by comparison in Fig.1c [15] with the data [7, 29]. Similarly, one cannot determine rigorously the threshold concentrations defining the mobility edge from experiments [16].

Finally, as a whole, the experimental results [15, 16] on properties of *n*- doped $SrTiO_3$ seem to be in stark contrast to the low temperature picture of a gas of the large polarons commonly used, for instance, for interpretation of the optical experiments (e.g.,[28]). The concept of the polaronic character of charge carriers in SrTiO should be revisited.

**ACKNOWLEDGMENTS**

The author thanks S. McGill and V. Dobrosavljevic for the stimulating discussions and Z. Fisk for reading the manuscript and his comments. I am grateful to H. J. Mard for creating the graphic material. The work is supported by the National High Magnetic Field Laboratory through NSF Grant No. DMR-1157490, the State of Florida and the U.S. Department of Energy.

## APPENDIX I

In [6] superconductivity of doper SrTiO₃ was ascribed to the "plasmon- polar optical phonon mechanism. The RPA expressions above account for the presence of the plasmon pole, however, for the Hamiltonian in form of Eq. (3) describing interaction of electrons with LO optical phonons its contribution to the superconducting pairing is negligible. Indeed, return to Eqs. (13, 14) of the main text. At $\omega_{nm}^2 > (\vec{v}_F \cdot \vec{q})^2$ in (14a):

$$\Gamma(q, \omega_{mn}) = -\frac{8\pi e^2}{q^2 \bar{\kappa}_\infty} \times \frac{\omega_{mn}^2}{\omega_{mn}^2 + \omega_{pl}^2}, \quad \text{(AIa)}$$

In (AIa) $\omega_{pl}^2 = 4\pi e^2 n_s / m$ is the square of the plasma frequency. On the axis of the *real* frequencies $\omega_{nm}^2 \to -(\varepsilon_n - \varepsilon_m)^2$ and $\Gamma(q, \omega_{mn})$ in Eq. (AIa) accepts the form

$$\Gamma(q, \omega_{mn}) \Rightarrow -\frac{8\pi e^2}{q^2 \bar{\kappa}_\infty} \times \frac{(\varepsilon_n - \varepsilon_m)^2}{\omega_{pl}^2 - (\varepsilon_n - \varepsilon_m)^2}. \quad \text{(AIb)}$$

That is, the plasma frequency pole indeed emerges in (AIb), however, for the plasma pole to play any role in the Cooper pairing its residue should be $\omega_{pl}^2$, not $(\varepsilon_n - \varepsilon_m)^2$. In the thermodynamic technique such pole is not singled out because with $\Gamma(q, \omega_{mn})$ in the form (AIa) summation over $\varepsilon_m$ does not contribute into the logarithmic singularity in Eqs. (9, 10). Correspondingly, the dependence on $\omega_{nm}$ in Eq. (14, 14a) can be omitted.

## APPENDIX II

A. The dimensionless $\bar{T}_{1,2,3}(x)$ in Figures 3a, b for the three bands are related to the corresponding temperatures of the Cooper instability as:

$$T_{C;1,2,3}(n_s) = (const)_{1,2,3} \times \frac{\gamma}{3\pi^3} \times \frac{\hbar^2}{m_i \bar{a}_B^2} \bar{T}_{1,2,3}(x)$$

B. Definitions of $\lambda_{1,2,3}(x)$ being as follows.

1) For the Lower band, Eq. (17):

$$\bar{T}_1(x) \equiv f_1(x) = x^2 \exp\left[-x / \ln(1+x)\right]$$

$$\lambda_1(x) = \frac{1}{x} \ln(1+x)$$

$$x = \pi p_F \bar{a}_B / \hbar \simeq 5.74(n_s \times 10^{-18})^{1/3}$$

2) In the Middle band, Eq. (19):

$$\bar{T}_2(x) = (x^2 - x_{c1}^2)\exp\left\{-\frac{1}{\lambda_2(x)}\right\}$$

$$\lambda_2(x) = \left(\frac{m_2}{m_1}\right)^{1/2} \times \frac{\ln[1 + F_2(x)]}{(x^2 - x_{c1}^2)^{1/2}}$$

$$F_2(x) = \left(\frac{m_2}{m_1}\right) \times \frac{x^2 - x_{c1}^2}{x + (m_2/m_1)^{3/2}(x^2 - x_{c1}^2)^{1/2}}$$

As a function of $x$, concentration of carriers in the second band $n_s = (3\pi^2 \hbar^3)^{-1}[p_{F1}^3 + p_{F2}^3]$ is given by the equation:

$$x^3 + (m_2/m_1)^{3/2}(x^2 - x_{c1}^2)^{3/2} \simeq 195 \times (n_s \times 10^{-18})$$

$$x > x_{c1} \simeq 6.1$$

3) For the Upper band, Eq.(21):

$$\bar{T}_3(x) = (x^2 - x_{c2}^2)\exp\left\{-\frac{1}{\lambda_3(x)}\right\}$$

$$\lambda_3(x) = \left(\frac{m_3}{m_1}\right)^{1/2} \times \frac{\ln[1 + F_3(x)]}{(x^2 - x_{c2}^2)^{1/2}}$$

$$F_3(x) = \left(\frac{m_3}{m_1}\right) \times \frac{x^2 - x_{c2}^2}{x + (m_2/m_1)^{3/2}(x^2 - x_{c1}^2)^{1/2} + (m_3/m_1)^{3/2}(x^2 - x_{c2}^2)^{1/2}}$$

The corresponding expression $n_s = (3\pi^2\hbar^3)^{-1}[p_{F1}^3 + p_{F2}^3 + p_{F3}^3]$ for the concentration of carriers in the third band as a function of $x$ is:

$$x^3 + (m_2/m_1)^{3/2}(x^2 - x_{c1}^2)^{3/2} + (m_3/m_1)^{3/2}(x^2 - x_{c2}^2)^{3/2} \simeq 195 \times (n_s \times 10^{-18})$$

$$x > x_{c2} \simeq 19.94$$